\documentclass[a4paper,floatfix,twocolumn,11pt,accepted=2022-06-08]{quantumarticle}
\pdfoutput=1
\usepackage[utf8]{inputenc}
\usepackage[english]{babel}
\usepackage[T1]{fontenc}
\usepackage{amsmath}
\usepackage{hyperref}
\usepackage{array}
\usepackage{multicol}
\usepackage{tabularx}
\usepackage{tikz}
\usepackage[numbers,sort&compress]{natbib}
\usepackage{floatrow}
\newfloatcommand{capbtabbox}{table}[][\FBwidth]

\usepackage{ulem}

\usepackage{siunitx}
\usepackage{blindtext}
\usepackage{amsmath}
\usepackage{amssymb}
\usepackage{physics}
\usepackage{mathtools}
\usepackage{dsfont}
\usepackage{xcolor}
\usepackage{float}
\usepackage{comment}
\usepackage{amsfonts}

\newcommand{\vek}[1]{\boldsymbol{#1}}

\DeclareMathOperator{\sinc}{sinc}

\newcommand{\dollar}{\$}
\newcommand{\cent}{\mbox{\textcent}}
\newcommand{\mymatrix}[2]{\left( \begin{array}{#1} #2 \end{array} \right)}
\newcommand{\myvector}[1]{\mymatrix{c}{#1}}
\newcommand{\mypar}[1]{\left( #1 \right)}

\begin{document}

\title{Quantum advantage using high-dimensional twisted photons as quantum finite automata}

\author{Stephen Z. D. Plachta}
\email{stephen.plachta@oeaw.ac.at}
\affiliation{Tampere University, Photonics Laboratory, Physics Unit, Tampere, FI-33720, Finland}
\affiliation{Institute for Quantum Optics and Quantum Information (IQOQI) Vienna, Austrian Academy of Sciences, Vienna, Austria}

\author{Markus Hiekkam\"aki}
\affiliation{Tampere University, Photonics Laboratory, Physics Unit, Tampere, FI-33720, Finland}

\author{Abuzer Yakary{\i}lmaz}
\email{abuzer.yakaryilmaz@lu.lv}
\affiliation{Center for Quantum Computer Science, Faculty of Computing, University of Latvia, R\={\i}ga, Latvia}
\affiliation{QWorld Association, Tallinn, Estonia, \url{https://qworld.net}}

\author{Robert Fickler}
\email{robert.fickler@tuni.fi}
\affiliation{Tampere University, Photonics Laboratory, Physics Unit, Tampere, FI-33720, Finland}

\begin{abstract}
Quantum finite automata (QFA) are basic computational devices that make binary decisions using quantum operations.
They are known to be exponentially memory efficient compared to their classical counterparts.
Here, we demonstrate an experimental implementation of multi-qubit QFAs using the orbital angular momentum (OAM) of single photons.
We implement different high-dimensional QFAs encoded on a single photon, where multiple qubits operate in parallel without the need for complicated multi-partite operations. 
Using two to eight OAM quantum states to implement up to four parallel qubits, we show that a high-dimensional QFA is able to detect the prime numbers 5 and 11 while outperforming classical finite automata in terms of the required memory.
Our work benefits from the ease of encoding, manipulating, and deciphering multi-qubit states encoded in the OAM degree of freedom of single photons, demonstrating the advantages structured photons provide for complex quantum information tasks.
\end{abstract}

\maketitle

Many scientific efforts are focused on controlling quantum systems, e.g., ions, atoms, electrons, photons, and others, that can be used to perform computational tasks more efficiently than classical computers. 
One of the simplest computational models is the finite automaton \cite{sipser}, which is a fundamental computational device that makes binary decisions (yes or no) using its finite memory (states) after reading a given input once, symbol by symbol.
More than two decades ago, it was shown that finite automata working on quantum mechanical principles, i.e., so-called quantum finite automata (QFA) using quantum states and quantum operations, can be exponentially more memory efficient than classical finite automata \cite{AF98}: for a given prime number $p$, QFAs can determine whether the length of any input is a multiple of $ p $ or not by using $ \mathcal{O}(\log p) $ states (with small decision errors), while any classical automaton requires at least $ p $ states \cite{AF98}.

In this article, we describe an experimental realization of a photonic QFA that implements an algorithm to decide whether the length of an input string is a multiple of a prime number.
We implement a high-dimensional QFA consisting of up to four 2-dimensional sub-automata that utilize the OAM modes of the high-dimensional state space of spatial light modes.
In our implementation, we show the quantum superiority of QFAs in terms of memory efficiency by distinguishing strings with lengths that are multiples of 5 and 11 from others using QFAs with only four and eight states, respectively.
We were able to experimentally implement these tasks with false decision probabilities below 10\% and 25\%, respectively.
In our implementation, we benefit from mature technologies that enable the flexible preparation and measurement of complex high-dimensional quantum states encoded in the transverse structure of single photons.
In contrast to earlier implementations, our implementation does not require controlled operations, allows a simple optical realization benefiting from the rotational symmetries of the spatial mode, and is realized along a single optical path. 
Hence, by using spatial modes, the operational state space can be scaled to higher dimensions without multi-partite entanglement or interferometric stability.
In theory, this state space can be arbitrarily large (being limited only by the aperture of the optical system).  

\section{Quantum Finite Automata}
\label{sec:QFA}
Before introducing the concept of QFA in more detail, we briefly discuss two important classical versions of finite automata \cite{sipser}.

\paragraph{Deterministic Finite Automata:} Maybe the simplest computational model is the so-called deterministic finite automaton (DFA) \cite{RS59}. 
A DFA can be used as a decider to determine, for example, whether a given input is in a recognized language or not, using a constant amount of memory, i.e., the capacity of memory does not change with the input size (length).
Formally, the memory of a DFA is represented by a finite set of states in which the automaton can be.\footnote{The number of different configurations that a DFA can be in during its computation is finite. At any step of its computation, a DFA is in exactly one of these configurations. Technically each configuration is called a state. For example, if the memory is formed by 10 bits, then the number of states is $ 2^{10} = 1024 $.}
At the beginning of the computation, it starts in a specified state called the initial state.
The computation is then performed by reading a given input (as a string) symbol by symbol from left to right. 
For each symbol, the state is updated with respect to its transition rules.\footnote{Each transition rule is represented as $ s \xrightarrow{\sigma} s' $, which means that when the automaton is in state $s$ and reads symbol $ \sigma $, the state is updated as $s'$. Note that for each pair of $s$ and $\sigma$, a transition rule must be defined. All transition rules of a DFA can be seen as its program.}
Each state is either an accepting or a rejecting state. 
Thus, at the end of the computation, the input is accepted or rejected if the final state is an accepting or a rejecting state, respectively. All accepted strings form a language, which is said to be recognized by this DFA.

While in general many different languages can be defined and recognized by DFAs, we use the following language family: for $n>1$, $ MOD_n = \{ a^j ~|~ j \mod n \equiv 0 \} $, where $ a^j $ denotes the concatenation of $ j $ $a$-symbols. 
In this case, the decision problem is whether the length of a given input string is a multiple of $ n $ or not. 
We can design a DFA $ D $ for $ MOD_n $ with $n$ states $ \{s_0,s_1,\ldots,s_{n-1}\} $. 
The computation of $ D $ starts in the initial state $ s_0 $. 
For each symbol $a$, the DFA performs a transition between the states according to
\[
    s_0 \xrightarrow{a} s_1 \xrightarrow{a} s_2 \xrightarrow{a} \cdots \xrightarrow{a} s_{n-1} \xrightarrow{a} s_0 
\]

After reading a string of $ n $ symbols, the system transitions back to its initial state $ s_0 $, if the length of the string is equal to a multiple of the number of states $n$. 
Thus, by setting $ s_0 $ to be the only accepting state, we have a DFA that recognizes the language $ MOD_n $ without any errors. 
It is easy to see that there is no DFA having less than $ n $ states for $ MOD_n $. 

\paragraph{Probabilistic Finite Automata:} Another version of a finite automaton is a probabilistic finite automaton (PFA) \cite{Rab63}, which is a generalization of a DFA as it can make more than one transition with probabilities that sum up to 1. 
The computation of a general $m$-state PFA, say $ P $, on a given input string $ x $ with $ l $ symbols (i.e., $ x = x[1]x[2]\cdots x[l] $) can be traced linearly. 
At any step, the PFA $ P $ is in a probability distribution of its states, which is represented by some $ m $-dimensional vector $ v = (p_1 ~~ p_2 ~~ \cdots ~~ p_m)^T $, where $p_j$ is the probability of being in the $ j $-th state. 
We call $v$ a state vector.\footnote{Different from a DFA, a PFA can be in more than one state with some probabilities summing up to 1. We use a vector to represent this probability distribution over the states, where the size of the vector is the number of states.}

For pre- and post-processing, $ P $ reads the special symbol $ \cent $  before the input and the symbol $ \dollar $ afterwards. 
At the beginning of the computation, the PFA $ P $ starts in the initial state vector $ v_0 $, where the entry corresponding to the initial state is 1 and the remaining entries are zeros.

For each symbol $x[j]$, there is a (left) stochastic transition matrix $A_{x[j]}$ (formed by the transition probabilities between the $m$ states when reading $x[j]$).
Remark that the transition matrix is the same for identical symbols.
After reading the whole input (including $\cent$ and $\dollar$), the final state vector is
\begin{equation}
    v_{f} = A_\dollar A_{x[l]} A_{x[l-1]} \cdots A_{x[1]} A_{\cent} v_0.
    \label{eq:PFA_final_state_vector}
\end{equation}
The probability with which the PFA $ P $ accepts $x$ is the summation of the entries of the final state vector $v_f$ corresponding to the accepting state(s).

Contrary to the case of a DFA, a PFA does not simply accept or reject the input. 
It assigns a probability in $ [0,1]$ to making its decision. 
Thus, the PFA $ P $ may make its decisions with errors. 
There are two basic decision modes for PFAs: unbounded-error and bounded-error. 

A language $ L $ is said to be recognized by PFA $ P $ with unbounded-error\footnote{In general, the threshold does not need to be $ \frac{1}{2} $. It can be any value in $ [0,1) $ called the cutpoint.} if and only if
\begin{itemize}
    \item any $x \in L$ is accepted by $ P $ with probability greater than $\frac{1}{2}$, and,
    \item any $ x \notin L $ is accepted by $ P $ with probability at most $\frac{1}{2}$.
\end{itemize}

A language $ L $ is said to be recognized by PFA $ P $ with error bound $ \varepsilon \in [0,1/2)  $ if and only if
\begin{itemize}
    \item any $x \in L$ is accepted by $ P $ with probability greater than $ 1-\epsilon$, and,
    \item any $ x \notin L $ is accepted by $ P $ with probability at most $\epsilon$.
\end{itemize}

In the latter case, by executing several copies of $ P $ in parallel and accepting the inputs by majority, the error bound can be arbitrarily close to zero,\footnote{As the number of copies does not depend on the input length, the obtained automaton is still a PFA.} thereby making the decision more reliable. 
It is common practice to fix the error bound as $ \frac{1}{3} $ (see for example page 397 of the standard textbook \cite{sipser} or the survey \cite{Watrous09}).

\paragraph{Quantum Finite Automata:} A QFA is the quantum counterpart of a PFA \cite{AY21}, since QFAs use the same framework due to the probabilistic nature of quantum physics. 
Here, we focus on the most restricted QFA model known, defined in \cite{MC00}. 
Any $m$-state QFA forms an $m$-dimensional Hilbert space ($\mathcal{H}^m$), and its state vectors are represented as the complex vectors $\ket{v}$ in $\mathcal{H}^m$. 
Further, the stochastic operators are replaced by unitary operators, and the final state is measured by projecting it onto the set of accepting states. 
The input is accepted if the measured state is in this subspace. 

Let $ Q $ be an arbitrary QFA with $ m $ basis states: $ N = \{\ket{q_1},\ldots,\ket{q_m}\} $. 
As in the case of a PFA, we define an input string $ x $ containing $ l $ symbols: $ x= x[1] x[2] \cdots x[l] $. 
The initial quantum state is $ \ket{v_0} = \ket{q_I} $, where $ q_I $ denotes the initial state. 
For each reading of the symbol $ x[j] $, the QFA performs the unitary transformation $ U_{x[j]} $ on the state. 
After reading the whole string, the final quantum state $\ket{v_f}$ is 
\begin{equation}
    \ket{v_f} = U_\dollar U_{x[l]} U_{x[l-1]} \cdots U_{x[1]} U_{\cent} \ket{v_0}.
\end{equation}
Thus, after a complete reading of the string $x$, a measurement of the final state $\ket{v_f}$ leads to an accepting probability of $ Q $, which is
\begin{equation}
    \label{eq:probability}
    \sum_{q_j \in N_a} | \braket{q_j}{v_f} |^2,
\end{equation}
where $N_a$ denotes the set of accepting states and $ N_a \subseteq N $.

Although the described QFA appears very similar to a PFA, the quantum nature of a QFA provides an exponential quantum advantage. Compared to bounded-error PFAs as well as DFAs, bounded-error QFAs can be exponentially more efficient in terms of memory (resources) \cite{AF98}. 
As mentioned before, any DFA recognizing the language $ MOD_n $ requires at least $ n $ states. 
For any prime number $p$, it is further known that any bounded-error PFA recognizing $ MOD_p $ also requires at least $ p $ states \cite{AF98}. 
On the other hand, bounded-error QFAs recognize the language $ MOD_p $ by using only $ \mathcal{O}(\log p) $ states \cite{AF98,AN09}. 
 
In the current work, we implement QFA algorithms for some $ MOD_p $ languages using less than $p$ quantum states.
Their functioning is described in detail in the following two sections, beginning with a 2-state QFA and expanding it to $2d$-state QFAs, where the quantum advantage becomes apparent.

\paragraph{2-state QFA:} When constructing a QFA for recognizing the language $MOD_p$ with a fixed prime number $p$, we first start with a simple 2-state QFA $ Q_2 $.
Its basis states are $ \ket{0} $ and $ \ket{1} $, i.e., a single qubit.
The state $ \ket{0} $ is the initial state as well as the only accepting state.
In the case of this 2-state QFA, the identity operator is applied when reading $ \cent $ and $ \dollar $.
The given input string $x$ consists of $n$ consecutive $a$-symbols, i.e., $ x = a^n $, following the notation given above.
For each symbol $a$, the state vector is rotated by an angle $ \theta = \frac{2\pi}{p} $ in the counter-clockwise direction on the real plane spanned by $ \{\ket{0},\ket{1}\} $.
Thus, a reading of the symbol $a$ leads to an evolution of the state with the unitary
\begin{equation}
\label{eq:Unitary_2StateQFA}
U_a = \mymatrix{rr}{ \cos \theta & -\sin \theta \\ \sin \theta & \cos \theta }.
\end{equation}
After reading the whole input, corresponding to $n$ operations $U_a$, the final quantum state is
\begin{equation}
    \ket{v_f} = \myvector{ \cos n \theta \\ \sin n \theta }.
    \label{eq:final_qu_state}
\end{equation}
Thus, the input is accepted with probability $ \cos^2 \mypar{ \frac{2 n \pi }{p} } $.
In this simple 2-state QFA $Q_2$, the accepting probability is always 1 when $ n $ is a multiple of $ p $.
For any other symbol, a low accepting probability is required to build a good decider.
However, the false acceptance probability of $Q_2$ can be up to $ \cos^2 \mypar{ \frac{\pi}{p} } $, i.e., the closest point of the final state vectors (for non-members) to the $ \ket{0} $-axis.
Hence, by choosing the rotation angle $\theta$ as defined above, a 2-state QFA $Q_2$ is able to recognize the input of length $p$ perfectly; however, nearly all other inputs, i.e., non-members of $MOD_p$, are erroneously accepted with a nonzero probability.

We note that, even though the mathematical description of $ Q_2 $ is simple and uses only real numbers, real quantum hardware uses complex-valued basis operators \cite{SY21A}.
Hence, a 2-state quantum system, i.e., a qubit, is commonly represented on a 3-dimensional Bloch sphere.
As a result, the rotations performed by the unitary operation $U_a$ can be implemented on a unit circle in an arbitrary plane of the Bloch sphere, a feature of quantum states which we apply in the experimental implementation described below.

\begin{figure*}[htb]
    \centering
    \includegraphics[width = 0.8\textwidth]{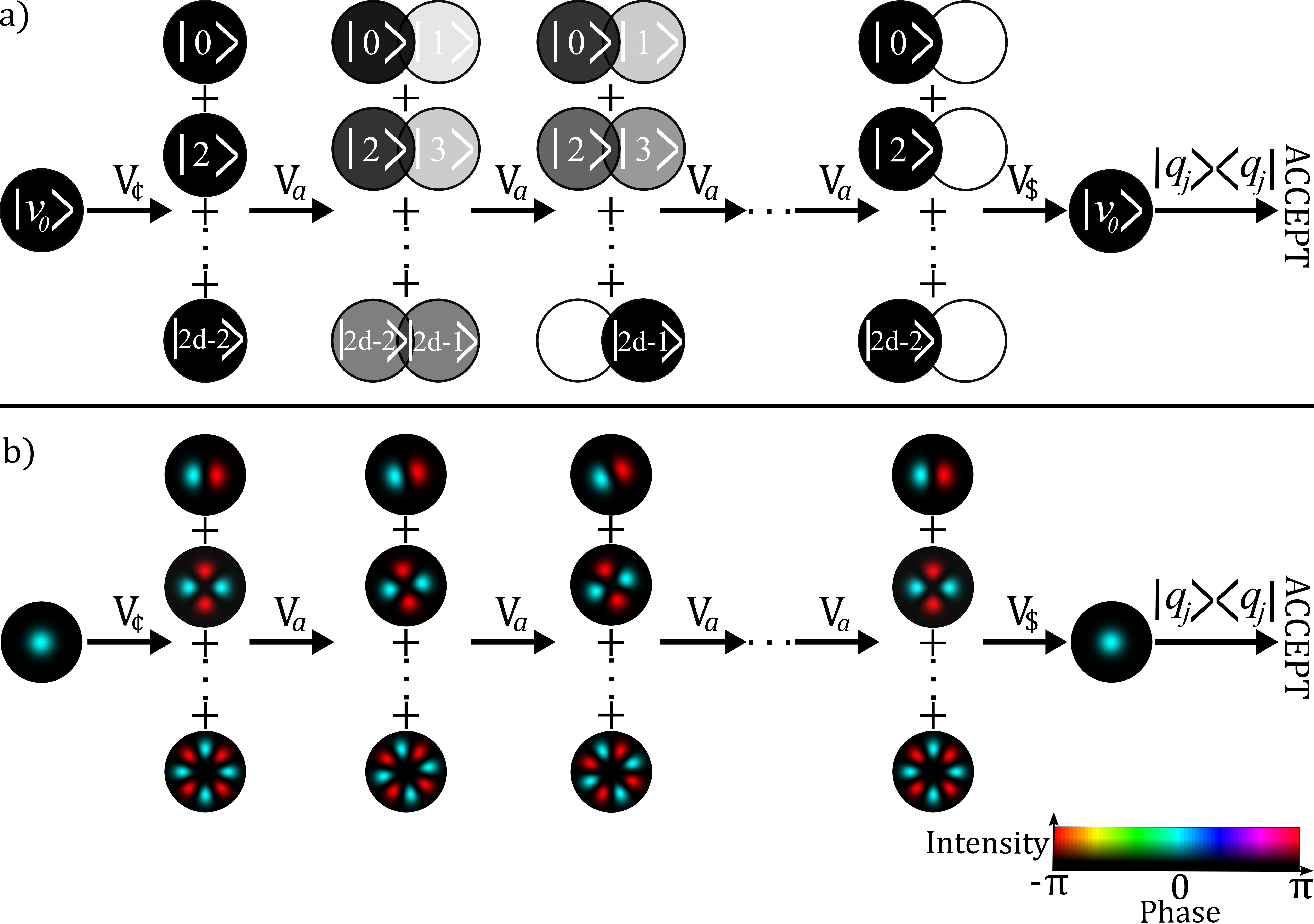}
    \caption{Conceptual graph describing \textbf{a)} an abstract $2d$-state QFA and \textbf{b)} the $2d$-state QFA implemented using photon OAM. In \textbf{a)} the input state $\ket{v_0}$ is first transformed into the initial superposition of multiple sub-automata $M_{k_j}$ through the unitary $V_{\cent}$, after which the states of these automata are evolved simultaneously through the unitary $V_a$. Note that each sub-automaton evolves at a different rate due to the differing values $k_j$. After $n$ operations of the unitary, all of the sub-automata have returned to their initial states, and an inverse operation $V_{\dollar} = V_{\cent}^{\dagger}$ is applied to return the superposition to the initial, accepting state. In \textbf{b)} this same process is described for an OAM QFA with a Gaussian initial state and OAM values $\ell = \{1,2,..,4\}$. Note also that the unitary $V_a$ is defined slightly differently in the case of the OAM QFA (see Eq.~\eqref{eq:multirot} and Eq.~\eqref{eq:multirot_petal}).
    }
    \label{fig:concept}
\end{figure*}

\paragraph{2\textit{d}-state QFA:} We now present a $ 2d $-state QFA $Q_{2d}$ that reduces the unwanted accepting probability for the non-members of $ MOD_p $, in comparison to a 2-state QFA, while still having less than $p$ states.
For any $ k \in \{1,\ldots,p-1\} $, we define a two-state QFA $ Q^k_2 $ by modifying $Q_2$ such that the value of $ \theta $ is set to $ \frac{ 2 k \pi }{p} $.
Again, each member of $ MOD_p $ is accepted by $ Q^k_2 $ with probability 1. Although the maximum accepting probability for a non-member is still $ \cos^2 \mypar{ \frac{\pi}{p} } $, the same non-member input string can have different accepting probabilities for $ Q^k_2 $'s with different values of $k$.
Thus, we can run a single high-dimensional QFA $Q_{2d}$, which includes several $ Q^k_2 $'s running in parallel, to reduce the error for the non-member strings.

To construct a $2d$-state QFA $ Q_{2d} $ such as the one depicted in Fig.~\ref{fig:concept}~a), we define $d$ different $k$ values, i.e., the set $ K = \{ k_1,\ldots,k_d \} $.
The basis states of $Q_{2d}$ are 
\begin{equation}
    \label{eq:basis_Mk}
    \{ \ket{0},\ket{1},\ket{2},\ket{3},\ldots,\ket{2d-2},\ket{2d-1} \}.
\end{equation}
As before, there is only one accepting state, which is the initial state $ \ket{v_0} $.
However, in contrast to $Q_2$, the reading of $ \cent $ leads to an application of the unitary operator $ V_{\cent} $, which transforms the initial state of the automaton $\ket{v_0}$ into the following superposition:
\begin{equation}
\label{eq:abstract_initial_state}
    V_{\cent}\ket{v_0} = \frac{1}{\sqrt{d}} \mypar{ \ket{0}+\ket{2}+\ket{4}+\cdots+\ket{2d-2} }.
\end{equation}
Here, the states $ \ket{2j-2} $ and $ \ket{2j-1} $ form a sub-QFA $Q^{k_j}_2$ for which a reading of the symbol $a$ corresponds to a rotation of the state by the angle $\theta_{k_j}=\frac{2 k_j \pi}{p}$, where $ 1 \leq j \leq k $.
The unitary matrix of each sub-QFA is defined as
\begin{equation}
\label{eq:Unitary_SubQFA}
U_{k_j} = \mymatrix{rr}{ \cos \theta_{k_j} & -\sin \theta_{k_j} \\ \sin \theta_{k_j} & \cos \theta_{k_j} }.
\end{equation}
Thus, a reading of a symbol $a$ causes a simultaneous rotation of all sub-QFAs, each rotating by a different angle $\theta_{k_j}$.
Hence, the total unitary state evolution $V_a$, when reading the symbol $ a $, is
\begin{equation}
\label{eq:multirot}
V_a = \left(\begin{array}{cc|cc|c|cc}
        &         &     0 &      0 &\cdots & 0 & 0\\
      \multicolumn{2}{c|}{\smash{\raisebox{.5\normalbaselineskip}{$U_{k_1}$}}}
                  & 0  &      0  &\cdots & 0 & 0\\
      \hline \\[-\normalbaselineskip]
    0   &   0  &   &  &\cdots & 0 & 0  \\
      0   &   0      & \multicolumn{2}{c|}{\smash{\raisebox{.5\normalbaselineskip}{$U_{k_2}$}}} &\cdots & 0 & 0\\ \hline
      \vdots & \vdots & \vdots & \vdots & \ddots & \vdots & \vdots \\ \hline
    0   &   0 & 0 & 0 &\cdots &   &     \\
      0   &   0  & 0 & 0  &\cdots   & \multicolumn{2}{c}{\smash{\raisebox{.5\normalbaselineskip}{$U_{k_d}$}}}   
    \end{array}\right).
\end{equation}
This means that, during the computation, $ Q_{2d} $ is in a superposition of $ d $ sub-QFAs $Q^{k_j}_2$, which implement different rotations $ \theta_{k_j} $ in parallel.
After reading the concluding symbol $ \dollar $, the high-dimensional QFA $ Q_{2d} $ applies the inverse of $ V_{\cent} $, i.e., $ V_{\dollar} = V_{\cent}^{\dagger} $.
If the input is in $ MOD_p $, then the final state will be $ \ket{v_0} $, the accepting state, and it will be accepted with probability 1.
More importantly, it was shown in \cite{AF98,AN09} that for each $ p $, there exists a unitary operator $ V_{a} $ and a set $ K $ with $ \frac{4}{\varepsilon} \log 2p $ elements such that any input not in $ MOD_p $ is accepted with probability no more than $\varepsilon $, where $ \varepsilon \in (0,1/2)$.
For a specific $ p $, one may check all possible $ k $ values numerically and identify the best ones.

\paragraph{Implementations of QFAs on real devices:}
The QFA algorithm for $ MOD_p $ is one of the first quantum algorithms with exponential advantage over classical methods of computing.
Its implementation for $ p=11 $ on superconducting quantum computers has been investigated in \cite{Kalis18,BSONY21,SY21A}.
However, the computation performed on the quantum computer became random immediately after reading a few symbols due to relatively large noise in the computation process.
An important reason for this deficiency was the usage of controlled operations when applying the unitary operators $ V_j $'s.
Such multi-qubit operations are very costly compared to other basic operations.

In \cite{MPC20,CMPCPO21}, a photonic implementation of a 2-state QFA was realized using the polarization degree of freedom of light. 
However, instead of implementing $n$ consecutive $ U_a $ operators, which would naturally lead to additional challenges and possible increased errors, a single operator $ U_{a}^n $, corresponding to the effective transformation of multiple $ U_a $ operators, was implemented.
Hence, the implementation simulates the computation of a QFA rather than performing it.

Another optical QFA has been demonstrated in \cite{tian2019experimental}. Using polarization and path encoding of quantum states, a QFA was implemented which was able to identify the prime numbers 3 from among the members of the set $\{3,4,5\}$ and 5 from $\{5,6,7\}$.
While in this implementation the correct accepting probability was very high ($>98~\%$), the implementation was limited due to the required consecutive waveplates and interferometric structure.

\subsection{QFA using photon OAM}
\label{sec:QFA_OAM}
Here, we overcome the deficiencies described above by experimentally implementing various $2d$-state QFAs using the transverse spatial modes of single photons.

We focus on encoding the quantum states in a specific set of transverse spatial modes, namely Laguerre-Gaussian (LG) modes \cite{andrews2012angular}. 
LG modes have become very popular as laboratory realizations of high-dimensional quantum states \cite{twist}.
They are solutions to the paraxial wave equation in cylindrical coordinates. 
Thus, they are described by two quantum numbers, i.e., the radial quantum number \cite{karimi2014radial} and the azimuthal quantum number \cite{mair2001entanglement}. 
The latter is commonly denoted by $\ell$, and it indicates that an LG mode has $\ell$ multiples of a $2\pi$ azimuthal phase gradient in its transverse structure, described by the helical phase factor $e^{i\ell\varphi}$.
This quantum number has attracted a lot of attention because it is connected to the number of quanta of OAM each photon carries \cite{Allen,mair2001entanglement}.
Over the last decades, it has been applied in a variety of different settings that span classical communications to quantum communications, computation, and simulation \cite{willner2015optical, twist, AndrewsBook, pinheiro2013vector}.
We implement the $2d$-state QFA using solely the OAM degree of freedom, i.e., photons with a radial quantum number of zero.

As shown in Fig.~\ref{fig:concept}~b), we construct the input QFA state via an operation $V_{\cent}$ that forms a superposition of $d$ two-dimensional sub-QFAs consisting of equally-balanced superpositions of two OAM modes with the same OAM but opposite sign.
Due to the lobes appearing in the transverse structures of these  superposition states, such light field modes are also called petal beams.
We use them to form the basis states of our QFA system, as in Eq.~\eqref{eq:basis_Mk}, which we can write using OAM modes $\ket{\ell}$ 
\begin{equation}
    \ket{p_{\ell_j}^\pm} = \frac{1}{\sqrt{2}}\left(\ket{\ell_j} \pm \ket{-\ell_j}\right).
    \label{eq:petal_basis}
\end{equation}
The positive petal modes ($p_{\ell_j}^+$) represent the even states ($\ket{2j-2}$), and the negative petal modes ($p_{\ell_j}^-$) represent the odd states ($\ket{2j-1}$).

To form an input state that corresponds to the state introduced in Eq.~\eqref{eq:abstract_initial_state}, we shape a single photon into a superposition of multiple positive petal modes:
\begin{equation}
    V_{\cent}\ket{v_0} = \frac{1}{\sqrt{d}}\left(\ket{p_{\ell_1}^+} + \ket{p_{\ell_2}^+} + \cdots  +\ket{p_{\ell_k}^+}\right).
    \label{eq:petals_initial_state}
\end{equation}
To operate on this state (read a character from the input string), we need a device that acts on the state via a unitary similar to that of Eq.~\eqref{eq:multirot}.
As shown in Fig.~\ref{fig:loop}, we select a Dove prism that rotates the field structure, thereby imprinting an OAM-dependent phase on the photons' azimuthal structure \cite{gonzalez2006dove}, though we note that schemes exist that are able to perform any unitary operation on structured photons \cite{brandt2020high}.
Through the Dove prism, each sub-QFA evolves similarly to Eq.~\eqref{eq:Unitary_SubQFA}, but it does so via a unitary which can be defined as
\begin{equation}
    \label{eq:petal_rot}
    U_a^{\ell_j} = \mymatrix{cc}{ \ \ \ \cos 2\ell_j\phi & -i \sin 2\ell_j\phi \\ -i\sin 2\ell_j\phi & \ \ \ \cos 2\ell_j\phi }
\end{equation}
in the petal mode basis.

Depending on the OAM $\pm\ell_j$ of the modes in the sub-QFA and the tilt angle $\phi$ of the Dove prism, the paraxial OAM modes propagating through the prism gain a phase of $e^{\mp i\ell_j2\phi}$, which corresponds to a rotation of the mode structure by  $2\phi$.\footnote{In the ray optics picture, there is a refraction through the $\ang{45}$ input face, a total internal reflection off the base that effectively imparts the $2\phi$ rotation, and a refraction through the $\ang{45}$ exit face that returns the beam to its original path.}
In other words, the Dove prism rotates each sub-QFA's state vector around its Bloch sphere axis defined by the two orthogonal OAM eigenstates $\ket{\ell_j}$ and $\ket{-\ell_j}$ \cite{padgett1999poincare}.

Analogous to the unitary introduced in Eq.~\eqref{eq:multirot}, the full unitary consists of multiple $2{\times}2$ blocks, which are now realized by the unitary defined in Eq.~\eqref{eq:petal_rot}, leading to
\begin{equation}
\label{eq:multirot_petal}
V_a = \left(\begin{array}{cc|cc|c|cc}
        &         &     0 &      0 &\cdots & 0 & 0\\
      \multicolumn{2}{c|}{\smash{\raisebox{.5\normalbaselineskip}{$U_a^{\ell_1}$}}}
                  & 0  &      0  &\cdots & 0 & 0\\
      \hline \\[-\normalbaselineskip]
    0   &   0  &   &  &\cdots & 0 & 0  \\
      0   &   0      & \multicolumn{2}{c|}{\smash{\raisebox{.5\normalbaselineskip}{$U_a^{\ell_2}$}}} &\cdots & 0 & 0\\ \hline
      \vdots & \vdots & \vdots & \vdots & \ddots & \vdots & \vdots \\ \hline
    0   &   0 & 0 & 0 &\cdots &   &     \\
      0   &   0  & 0 & 0  &\cdots   & \multicolumn{2}{c}{\smash{\raisebox{.5\normalbaselineskip}{$U_a^{\ell_k}$}}}   
    \end{array}\right).
\end{equation}
After $n$ operations of $V_a$ on $V_{\cent}\ket{v_0}$, the accepting probability corresponding to Eq.~\ref{eq:probability} is described by
\begin{equation}
    \abs{\bra{v_0}V_{\dollar} V_a^n V_{\cent}\ket{v_0}}^2 = \frac{1}{d^2}
    \left(
    \sum_{\ell_j=k_1}^{k_d}\text{cos}(2 n \ell_j \phi)
    \right)^2.
    \label{eq:acc_prob_QFA}
\end{equation}
\par

To generate a quantum decider for $MOD_p$ language with this photonic QFA, we link the tilt angle of the Dove prism $\phi$ to the value $p$ of the $MOD_p$ language.
By comparing Eq.~\eqref{eq:petal_rot} and Eq.~\eqref{eq:Unitary_SubQFA}, it is clear that we need to choose the Dove prism angle $\phi = \pi/p$ or $\phi = \pi/(2p)$. 
The latter of these can be chosen only when exclusively even or exclusively odd OAM values are used, since under these conditions,  all of the sub-QFAs return back to their initial state, with the same phase, when $2p\phi = \pi$.
In either of these cases, the accepting probability of Eq.~\eqref{eq:acc_prob_QFA} is 1 when $n=p$ (or an integer multiple of $p$).

It should be noted that a Dove prism also affects the polarization of the photon in addition to rotating the transverse structure \cite{gonzalez2006dove, padgett1999dove}.
In the present work, we eliminate any polarization change by adding additional waveplates (see Fig. \ref{fig:loop}), so we omit its detailed description here.
However, adding the polarization degree of freedom would add more states into the operational state space of a photonic QFA. 
Thus, it could be harnessed in future implementations to expand the underlying Hilbert space.

\section{Experimental Implementation}
\label{sec:exp}
To implement the high-dimensional QFA described above, we need a source of single photons, a device for preparing and measuring the state, and a method to implement consecutive unitary operations $V_a$.

As the single-photon source, we used a type 0 spontaneous parametric down-conversion (SPDC) process to generate pairs of photons at $810$~nm by pumping a nonlinear crystal (periodically poled potassium titanyl phosphate or ppKTP) using a $405$~nm laser. 
After the crystal, the pump wavelength is filtered out using bandpass filters, and the two photons of each pair are coupled into separate single mode fibers (SMF).
From the SMFs, one photon acts as a heralding trigger and goes straight to a single-photon detector, while the second photon (signal) is sent through the QFA system, after which it is registered by a second identical detector.
Using a coincidence counter, we post-select on photons from the same SPDC process using a coincidence window of around 1~ns.
A trigger photon thereby heralds the presence of its signal mate leading to a stream of single photons \cite{Mandel}.
We verified the high quality of the source by calculating a $g^{(2)}(0)$\nobreakdash-value of $0.022 \pm 0.003$ from coincidence measurements, following the process described in \cite{bouchard2018experimental}.
More details on the photon source can be found in Appendix~\ref{sec:det_exp}. 

\begin{figure}[htb]
    \centering
    \includegraphics[width=\textwidth]{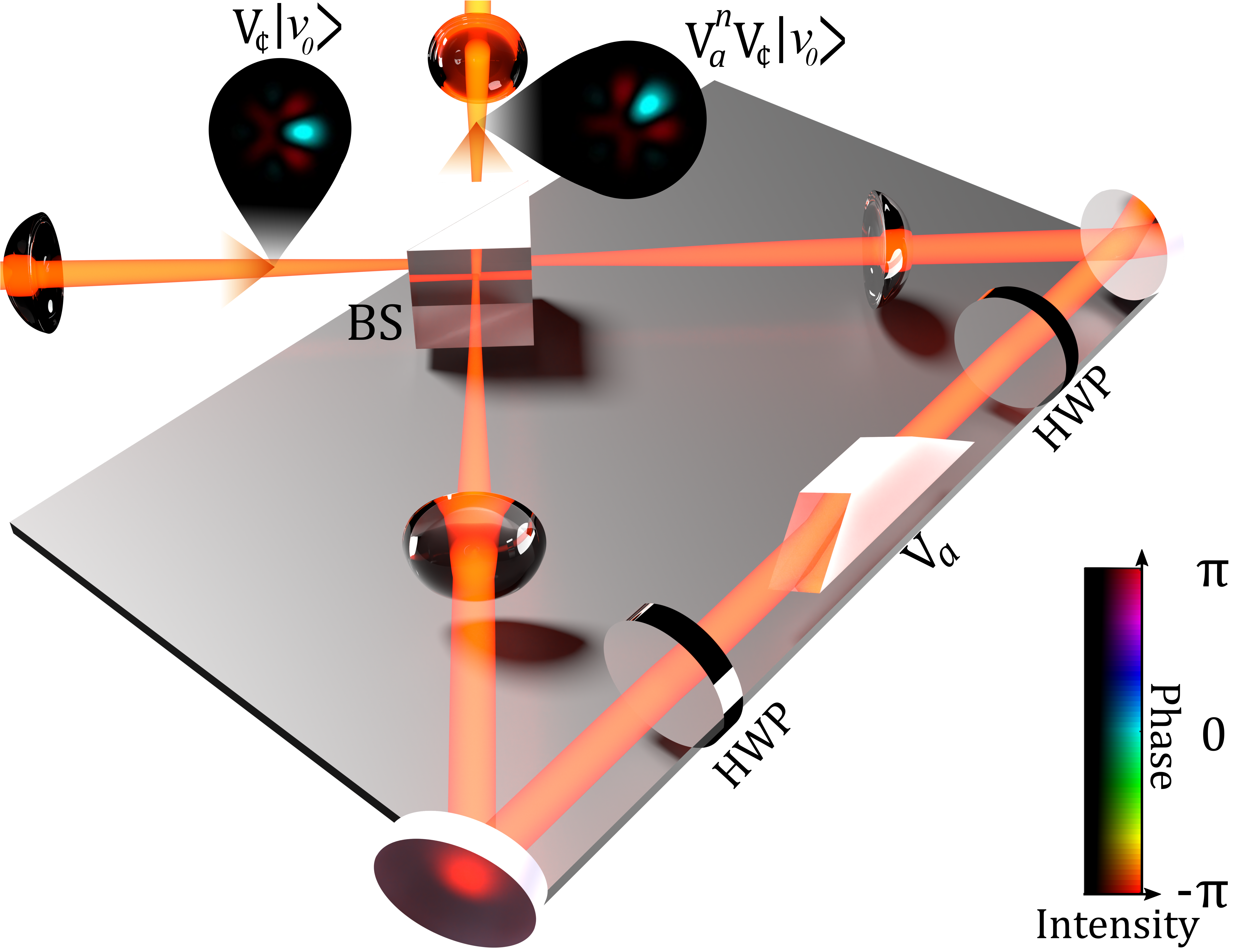}
    \caption{A conceptual render of the loop that performs consecutive unitary operations $V_a$ on the input state. 
    A photon in a superposition $V_{\cent}\ket{v_0}$ is sent into the loop from the left.
    In the loop, the light field's polarization is aligned with one of the sides of the Dove prism using a half-wave plate (HWP).
    This allows a Dove prism to rotate the transverse structure of the photon while its polarization remains unaffected \cite{padgett1999dove}.
    After the Dove prism, a second HWP rotates the polarization of the field back to a horizontal position, which is the polarization ensuring the most efficient operation of the SLM.
    A 50:50 (or 70:30) beamsplitter (BS) is used to keep the photons in the loop $50~\%$ ($70~\%$) of the time, allowing  multiple unitary operations to be performed on a photon.
    The four lenses are used to construct a continuous imaging system that removes the free-space diffraction during propagation.}
    \label{fig:loop}
\end{figure}

To encode the quantum state of Eq.~\eqref{eq:petals_initial_state} onto a signal photon, we employ a holographic technique using a spatial light modulator (SLM) that shapes the transverse spatial mode through amplitude and phase modulation \cite{Bolduc}.
In this technique, the desired complex structure is carved out of a large Gaussian beam using a single phase-only hologram (see Appendix~\ref{sec:Gauss_corr} for more information).
Despite being a lossy process, this state encoding can be seen as the transformation $V_{\cent}$ described earlier.
However, we note that these losses could be overcome using other, unitary encoding schemes \cite{hiekkamaki2019near}.

After this initialization, as shown in Fig.~\ref{fig:loop}, the signal photon enters an optical loop comprised of a beamsplitter and a Dove prism that performs consecutive unitary operations $V_a$ on the QFA state.
Since the Dove prism acts multiple times on a photon that stays in the loop, its rotational alignment requires a very high precision, achieved via a process detailed in Appendix~\ref{sec:Dove_angle}.
The number of operations performed on a photon depends on the probability it has of leaving the loop after $n$ round-trips, which is defined by the splitting ratio $R$:$T$ (reflectance and transmittance) of the beamsplitter.
Since each loop applies the unitary once to the QFA state, each trip through the loop corresponds to the reading of one symbol from the input string.
We note that, while in this first proof-of-principle implementation, the beamsplitter is probabilistically ``deciding'' the number of unitary operations $V_a$ applied to the photon, in general an active fast switching could be used in the future to implement our scheme in a controlled way.
The photon is reimaged by lenses inside the loop, which build one continuous imaging system of consecutive $4f$ systems.
The number of mirrors and focal points are chosen such that each operation rotates the field in the same direction while ensuring that no unintended flipping of the transverse structure occurs.
Finally, half-wave plates are placed before and after the Dove prism to eliminate the prism's effect on the polarization vector \cite{padgett1999dove} by returning it back to horizontal after each loop.
This is done to maintain optimal efficiency of the polarization-sensitive SLM used to measure the state of the photon after it leaves the loop.

In order to perform a measurement of the photon's transverse spatial structure, we use the same holographic technique as in the mode generation, which is now used to realize a filter for the accepting state.
We display the measurement hologram in a separate region of the same SLM screen.
Analogous to the generation process, despite the lossy nature of the scheme, the holographic modulation resembles the final unitary transformation $V_{\dollar}$.
The filter is realized by the measurement hologram in combination with a subsequent coupling into an SMF \cite{bouchard2018measuring} (see Appendix~\ref{sec:Gauss_corr} for more information).
If the photon's state and, thus, the incoming field structure after $n$ loops matches perfectly with the filtering hologram, then the photon's transverse structure is flattened, and it couples into the SMF with the highest possible efficiency.
In these cases, the accepting probability is 1.
A detailed description and sketch of the full setup is shown in Appendix~\ref{sec:det_exp}.

Because the photon undergoes $n$ operations probabilistically and experiences loop-dependent losses due to the optics, we must normalize the $n^{\text{th}}$ accepting probability of a structured photon.
We do so by gauging the efficiency of our system using a Gaussian mode (rotationally invariant structure) under the same experimental parameters.
In each QFA implementation, we first propagate a Gaussian state through the whole system and record the rate of the heralded single photons per loop.
As these values directly correspond to the loop-dependent efficiency of our setup, we use them to re-normalize the count rates of the complex QFA states to obtain the appropriate acceptance probabilities.
We accounted for accidental coincidences by removing them from the data during analysis.
Both procedures are described in more detail in Appendix~\ref{sec:loops_acc}.

\section{Results}
\label{sec:results}
We first report on a set of experiments where we implemented four different 2-state QFAs, which characterize our setup and demonstrate the flexibility of our approach.

\begin{figure}[t]
    \centering
    \includegraphics[trim={2.7cm 2.6cm 2.7cm 1.5cm},clip,width=\textwidth]{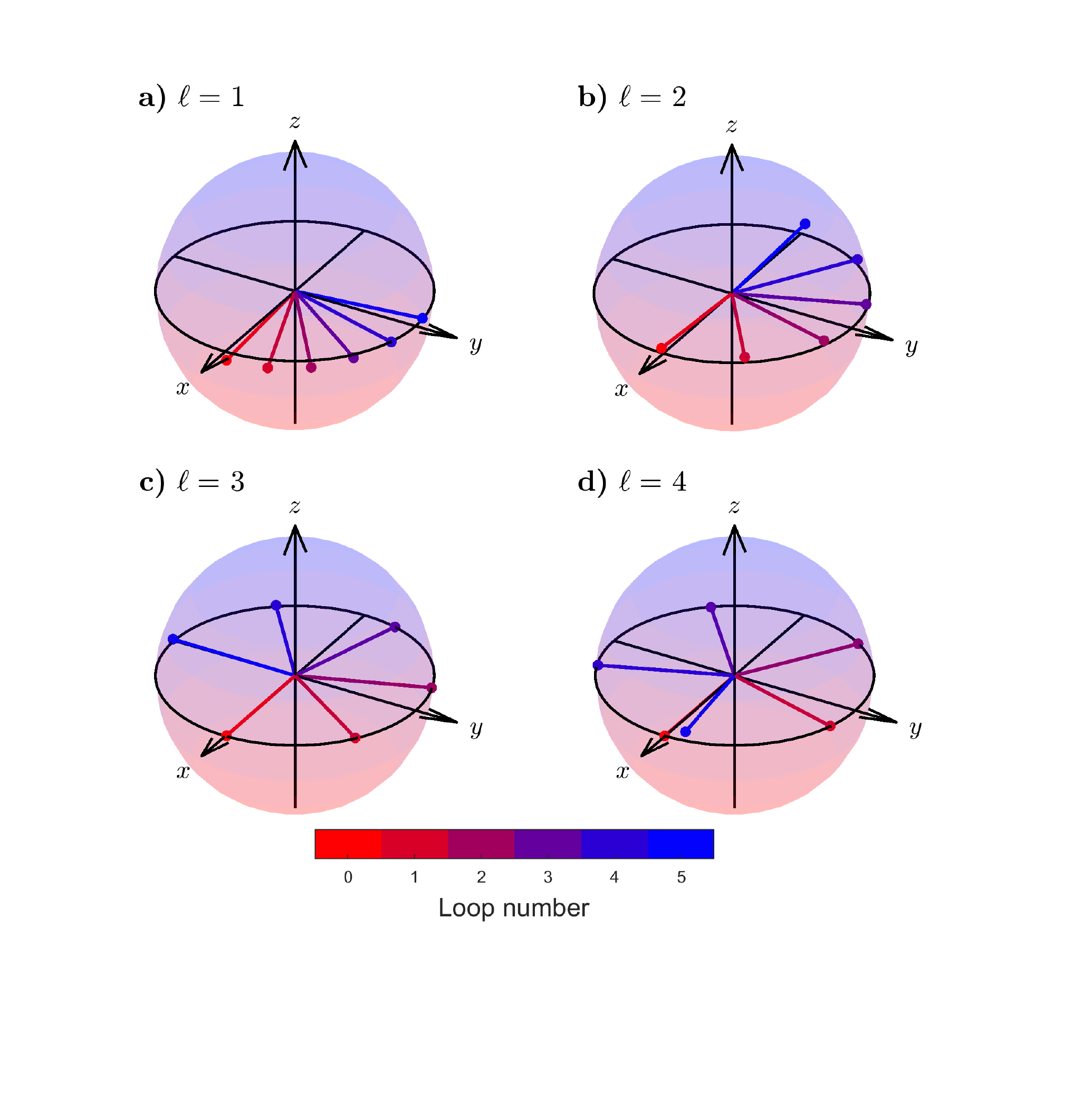}
    \caption{Quantum state tomography of 2-state QFA on the Bloch sphere with the $z$,$x$,and $y$ axes formed by the states $\{\ket{\ell},\ket{-\ell}\}$, $\{\frac{1}{\sqrt{2}}(\ket{\ell}+\ket{-\ell}), \frac{1}{\sqrt{2}}(\ket{\ell}-\ket{-\ell})\}$, and $\{\frac{1}{\sqrt{2}}(\ket{\ell}+i\ket{-\ell}), \frac{1}{\sqrt{2}}(\ket{\ell}-i\ket{-\ell})\}$, respectively. With the Dove prism at an angle of $\ang{4.5}$, the state vector rotates by $2\ell\times\ang{9}$ around the $z$-axis each time the photon passes through the loop. The total numbers of heralded photons for all state reconstructions per Bloch sphere were \textbf{a)} $\num{3.99e7}$ for $\ell=1$, \textbf{b)} $\num{2.67e7}$ for $\ell=2$, \textbf{c)} $\num{2.85e7}$ for $\ell=3$, and \textbf{d)} $\num{1.79e7}$ for $\ell=4$. }
    \label{fig:QST}
\end{figure}

As outlined in Sec.~\ref{sec:QFA}, a QFA system is described by multiple subsequent unitary operations performed on a quantum system.
In order to characterize these unitary evolutions, we performed full quantum state tomography (QST) on the state after zero to five unitary operations with the Dove prism set to an angle $\phi=\ang{4.5}$.
We used a $50$:$50$ beamsplitter for probabilistic in- and out-coupling to the loop.
Although the Dove prism angle was fixed, we used its OAM-dependent operation to implement different QFAs for OAM values $\ell = \{1,2,3,4\}$.
For each of these realizations, we can display its state evolution on its corresponding Bloch sphere, where the two poles correspond to the two OAM eigenstates of opposite sign \cite{padgett1999poincare}.

As shown in Fig.~\ref{fig:QST}~a)-d), the measured state vectors, calculated by direct inversion \cite{Schmied2016QST,Chithrabhanu2015generalized}, evolve in accordance with Eq.~\eqref{eq:petal_rot}: the states rotate at increased rates for higher-order OAM.
After five unitary operations, the final state vector has rotated by $\ell \times \ang{90}$.
Hence, using this setting of the Dove prism and an OAM value of $\ell=4$, we see that the final state matches the initial state after the fifth operation.
As such, it can be seen as a 2-state QFA recognizing the $MOD_5$ language (for more information on the accepting probabilities of the QFA, see Appendix~\ref{sec:QST_app}).

Note that the state vectors deviate slightly from the theoretically expected equally balanced superpositions which reside on the complex unit circle in the $xy$-plane, i.e., the equator.
The deviations are due to experimental imperfections in the imaging system, as well as the state generation and measurement methods, which cause small aberrations and slight misalignments.

\begin{figure}[t]
    \centering
    \includegraphics[width=\textwidth]{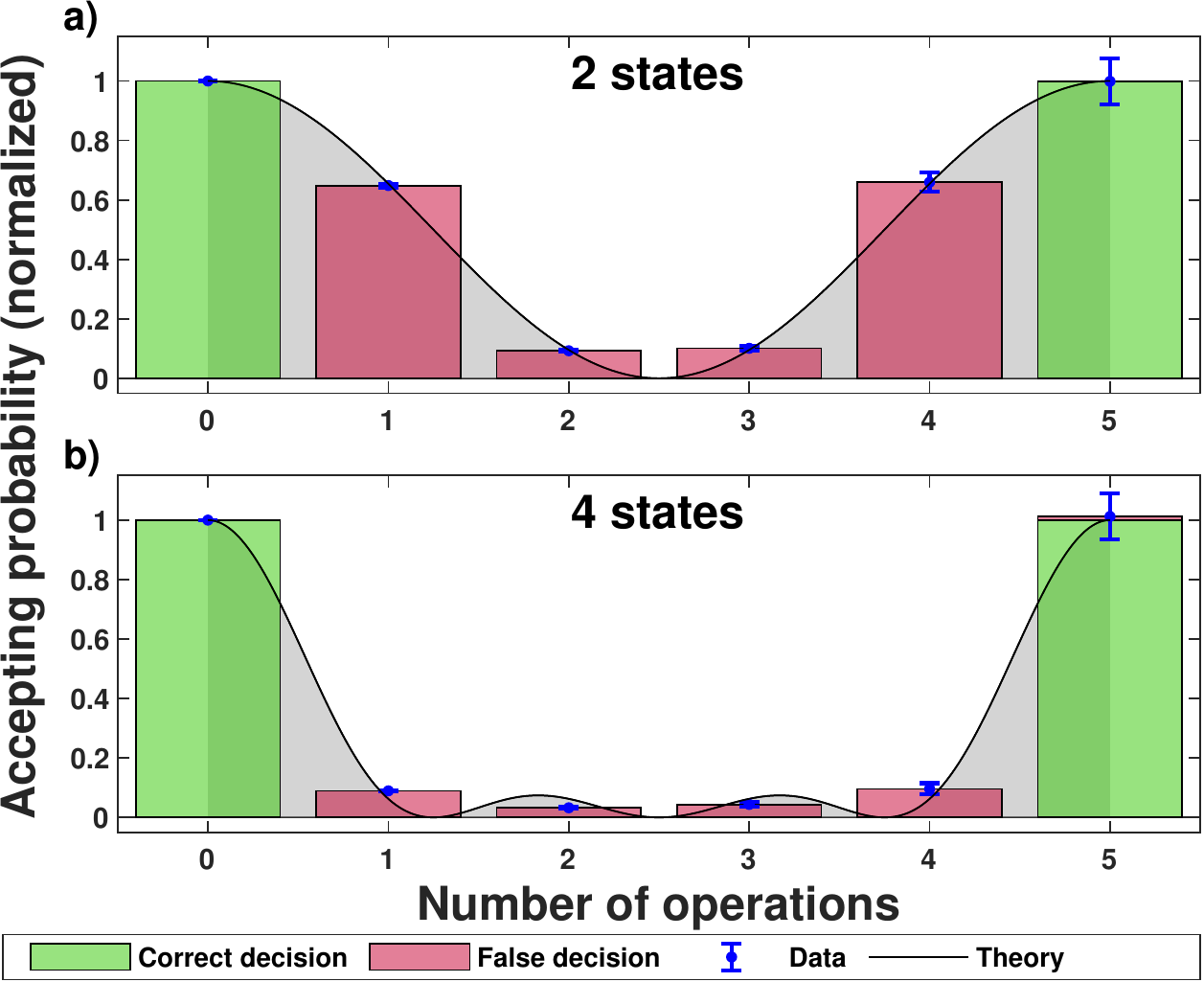}
    \caption{Decision probabilities for QFAs detecting strings belonging to $MOD_5$. \textbf{a)} A $2$-state QFA implemented with the OAM value of $\ell=1$. The data (blue points) and error bars ($\pm1$ standard deviation) were obtained from $152$ repetitions of the experiment with around $\num{7.89e4}$ heralded photons per run. \textbf{b)} A $4$-state QFA using two sub-QFAs of OAM values $\ell=\{1,3\}$. Here, the data and error bars were obtained from $58$ repetitions of the experiment with around $\num{4.07e4}$ heralded photons per run. The quantum advantage of the $2d$-state QFA in terms of memory efficiency is apparent, as only $4$ states are used, while a classical DFA requires $5$ states. The solid line shows the expected theoretical detection probability for a continuously-rotating state, calculated by Eq.~\eqref{eq:acc_prob_QFA}.}
    \label{fig:QFA_data1}
\end{figure}

We next report on our realization of a functioning 2-state QFA that recognized the $MOD_5$ language using the petal mode with $\ell = 1$.
Taking into account Eq.~\eqref{eq:petal_rot}, we find that with the Dove prism set to $\phi=\ang{18}$, the initial state encoded with this lowest OAM order returns to itself (up to a global phase) after $p=5$ loops.
As shown in Fig.~\ref{fig:QFA_data1}~a), a 2-state QFA implemented with these settings results in a very good acceptance probability, after 5 operations, of ${\sim}100~\%$ ($99.89\pm 7.71~\%$).
Note that the probability can go above 100~\% due to the normalization procedure (see Appendix~\ref{sec:loops_acc}).
However, as expected from a 2-state QFA, the false acceptance probabilities for some non-members of $MOD_5$ are beyond the reasonable error bounds mentioned in Sec.~\ref{sec:QFA}, e.g., as high as $64.84\pm 0.66~\%$ and $65.97\pm 3.19~\%$ for inputs of length 1 and 4, respectively.
Though these values correspond well to the maximum false acceptance of $\cos^2 \mypar{ \frac{\pi}{p} }$ discussed for $Q_2$, which is approximately $65.45~\%$ for $p=5$, such a QFA cannot be considered a properly working finite automaton.

In order to reduce the false acceptance probability, we implemented a $2d$-state QFA with $d=2$.
We realized this 4-state QFA by superposing the two 2-state sub-automata with OAM values $\ell = \{1,3\}$ according to Eq.~\eqref{eq:petals_initial_state}.
With a photon in such a state, the Dove prism acts on each sub-QFA in parallel according to the operation described by Eq.~\eqref{eq:multirot_petal}.
As shown in Fig.~\ref{fig:QFA_data1}~b), after five operations the implemented QFA again results in a near perfect acceptance probability ($101.22\pm 7.82~\%$), thus recognizing the $MOD_5$ language.
More importantly, the false acceptance probability for input strings not in $MOD_5$ was lowered to $9.62\pm 1.89~\%$ (loop 4) or less, which is well below the $\frac{1}{3}$ error bound and shows a significant improvement in performance.
The results demonstrate the quantum advantage of a QFA in terms of memory efficiency, as only four states are enough to recognize the $MOD_p$ language with $p=5$.
To illustrate the ease with which our system can be scaled to larger state spaces and how this affects the false decision probabilities, we include data for 6-state and 8-state QFAs recognizing $MOD_5$ in Appendix~\ref{sec:high-d}.
However, we note that these implementations do not demonstrate a quantum advantage.

To implement QFAs for larger prime numbers, we increased the number of detectable operations to larger values by modifying the beamsplitter of our system.
The probability of a photon leaving the loop after $n$ cycles ($n~{\in}~\mathbb{N}^+$) can be calculated from $P(n)=T^2 R^{n-1}$, where $T$ and $R$ are the transmittance and reflectance, respectively.
From this equation one can see that an unbalanced beam splitting ratio $R$:$T$ can increase the probability for larger $n$ while reducing it for the first few cycles.
Thus, for a certain level of background noise, an unbalanced beamsplitter allows us to get a usable signal from a greater number of loops.
We find that a $70$:$30$ beamsplitter results in fewer counts than a $50$:$50$ beamsplitter for loops 1-4 but increases the count rates for $n \geq 5$, and we were able to detect photons that cycled up to eleven times through the loop.

\begin{figure}[t]
    \centering
    \includegraphics[width=\textwidth]{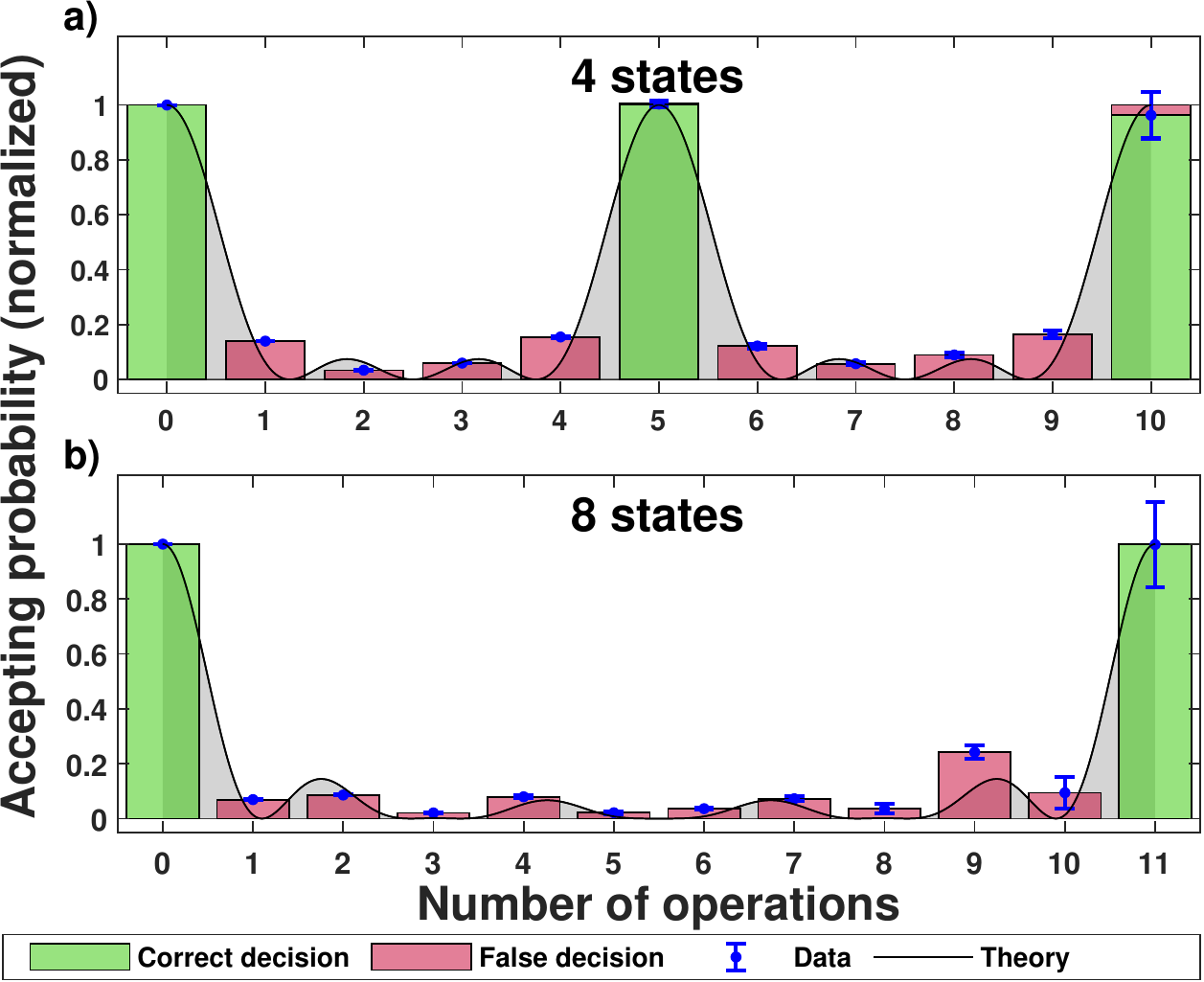}
    \caption{Decision probabilities for QFAs detecting strings belonging to $MOD_p$ with $p=5$ or $p=11$. \textbf{a)} The same $4$-state QFA ($\ell=\{1,3\})$ as in Fig.~\ref{fig:QFA_data1}~b), but now the operation was performed $10$ times, i.e., $2$ full cycles of the QFA. The data (blue points) and errors bars ($\pm1$ standard deviation) were obtained from $40$ repetitions of the experiment with around $\num{1.04e7}$ heralded photons per run. \textbf{b)} An $8$-state QFA with $\ell=\{1,2,3,4\}$ demonstrates an exponential quantum advantage in terms of memory efficiency for $p=11$. Here, the data and error bars were obtained from $50$ repetitions of the experiment with around $\num{1.70e6}$ heralded photons per run. The solid line shows the expected theoretical detection probability for a continuously-rotating state, calculated by Eq.~\eqref{eq:acc_prob_QFA}.}
    \label{fig:QFA_2cycle}
\end{figure}

As shown in Fig.~\ref{fig:QFA_2cycle}~a), with the same Dove prism angle ($\phi=\ang{18}$), we were able to detect photons that performed two full cycles of the 4-state QFA ($\ell = \{1,3\}$) by going through the optical loop ten times, which corresponds to reading a string of length 10.
The false acceptance was kept low, to maximally $16.45\pm 1.44~\%$ (loop 9), while the state was accepted after 5 and 10 loops with near-perfect probabilities of $100.33\pm 1.11~\%$ and $96.26\pm 8.45~\%$, respectively.

In the last set of experiments, we implemented a high-dimensional QFA showing a quantum advantage for the $MOD_{11}$ language.
With the Dove prism set to an angle of $\phi=\ang{360}/11 \approx \ang{16.36}$, we formed an 8-state QFA by superposing the four 2-state sub-QFAs with OAM values $\ell = \{1,2,3,4\}$.
As shown in Fig.~\ref{fig:QFA_2cycle}~b), we obtained a decision error of $24.18\pm 2.39~\%$ (loop 9) or less, while the system accepted the correct string of length 11 with a probability of $99.84\pm 15.61~\%$.
As before, the results nicely match the theoretical probabilities of Eq.~\eqref{eq:acc_prob_QFA} and demonstrate an exponential quantum advantage compared to classical FAs, which require eleven states.

\section{Discussion}
\label{sec:disc}
In conclusion, using the OAM degree of freedom of photons allowed us to demonstrate the superiority of a QFA system over classical finite automata when recognizing $MOD_p$ languages up to $p=11$.
Using the rotation of a photon's transverse structure, e.g., induced by a Dove prism, and higher order OAM states, we were able to perform a well-defined unitary operation acting simultaneously on multiple qubits with a single optical device.
Compared to earlier experimental QFA implementations, our result is not only the first to encode multiple qubits onto a single quantum system, but it also demonstrates some of the advantages OAM states offer in terms of simplicity and scaling for photonic quantum information processing.
In principle, the size of the underlying Hilbert space is only limited by the increasing size of the photon's structure and the aperture of the system.
As OAM states of up to 10010 have already been demonstrated \cite{fickler2016quantum}, QFAs with hundreds or more parallel, independent qubits could be envisioned.
Furthermore, even for such a large state space, the scheme introduced here does not require any controlled operations between multiple quantum systems since only high-dimensional operations on a single quantum system are required. 
However, increasing the dimensionality of the quantum states also requires us to increase the overall efficiency of the computation.
Here, a straightforward improvement would be to reduce the losses appearing in the pre- and post-processing operations $V_{\cent}$ and $ V_{\dollar} $, for which lossless operations are known \cite{hiekkamaki2019near}.
However, the main drawback of our current implementation in terms of efficiency is the probabilistic reading of symbols from the input string, which is done by keeping the photon in the loop using a passive beamsplitter.
Future implementations can be improved by including an active switch to control how many symbols are read, e.g., by fast polarization switching and a polarizing beamsplitter.
Thus, it is possible to implement the presented scheme in a fully controlled and lossless way, thereby showing the promise spatial structuring of photons offers.

\section*{Acknowledgments} 
The authors thank Rafael F. Barros for fruitful discussions and suggesting the unbalanced beam splitter. 
SZDP, MH, and RF acknowledge the support of the Academy of Finland through the Competitive Funding to Strengthen University Research Profiles (decision 301820), (Grant No. 308596), and the Photonics Research and Innovation Flagship (PREIN - decision 320165). 
MH acknowledges support from the Doctoral School of Tampere University and the Magnus Ehrnrooth foundation through its graduate student scholarship.
AY was partially supported by the ERDF project Nr. 1.1.1.5/19/A/005 ``Quantum computers with constant memory'' and the project ``Quantum algorithms: from complexity theory to experiment'' funded under ERDF programme 1.1.1.5.
RF acknowledges support from the Academy of Finland through the Academy Research Fellowship (Decision 332399).

\bibliographystyle{plainnat}
\bibliography{Biblio}

\onecolumn\newpage
\appendix
\section{Detailed experimental scheme}
\label{sec:det_exp}
\begin{figure*}[h]
    \centering
    \includegraphics[width = \textwidth]{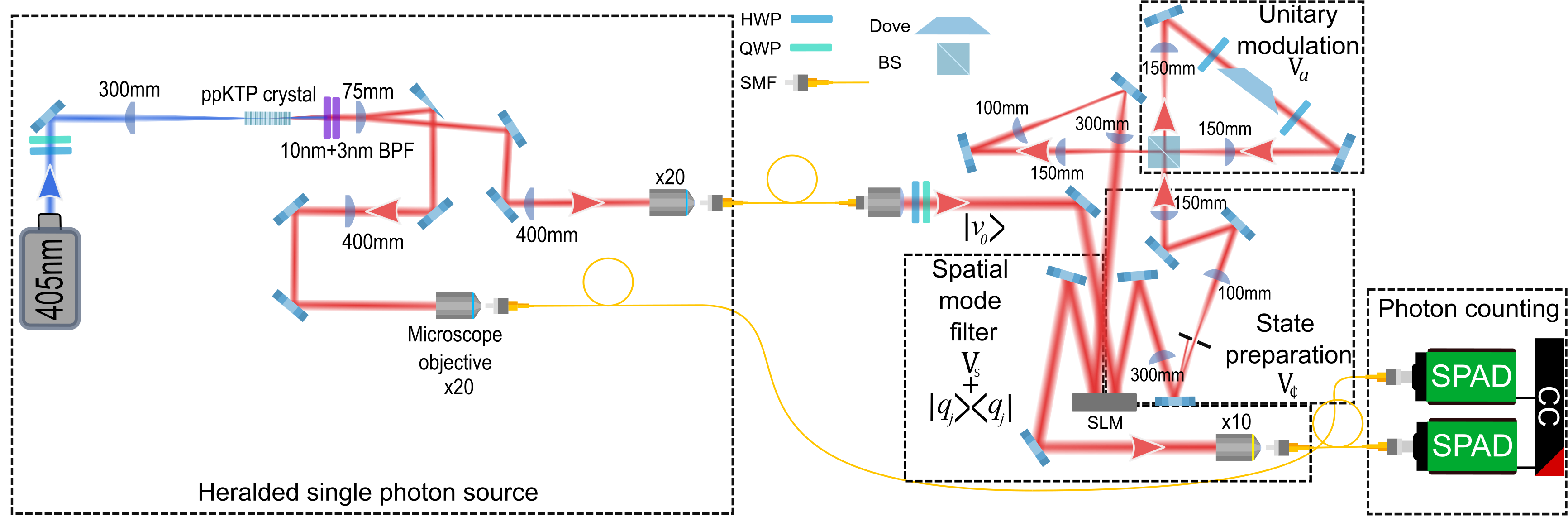}
    \caption{A detailed drawing of the experimental setup. The focal length of each lens is given in the figure, along with the magnifications of the used microscope objectives. All SMF's were placed on coupling stages (xyz).
    The HWP's and quarter wave plates (QWP) are used to facilitate optimum efficiency of the polarization dependent SPDC process and the SLM.}
    \label{fig:detailed_setup}
\end{figure*}

As depicted in the experimental setup of Fig.~\ref{fig:detailed_setup}, the photon pairs are created in an SPDC source.
The source uses a 12~mm long, type 0 periodically poled potassium titanyl phosphate (ppKTP) nonlinear crystal to convert photons of the 405~nm continuous-wave pump laser into degenerate photon pairs at 810~nm.
The slightly astigmatic focus of the pump beam had an approximately 67~$\mu$m Gaussian beam waist in the crystal.
The down-converted photons are made degenerate by tuning the phase matching of the crystal by controlling its temperature. 
The pump is then removed using bandpass filters (BPF) that also filter the down-conversion down to a bandwidth of 3~nm. 
The photons are then separated from each other using their momentum anti-correlations, i.e., we use a lens to perform an optical Fourier transform of the down-conversion and split the photons using a D-shaped mirror placed in the middle of the beam, one focal distance away from the lens.
Both photons are then coupled into single mode fibers (SMF) on coupling stages (xyz) using 400~mm lenses and microscope objectives.
From the SMFs, the heralding photon is sent straight to a single photon avalanche photo diode (SPAD: laser components COUNT-T), and the signal photon is sent to the QFA system.

In the QFA system, the photon coming out of the SMF is in the eigenmode of the SMF (Gaussian) and is first collimated and sent through waveplates which align its polarization along the plane of the image.
The photon's spatial mode is then structured using the Gaussian-corrected mode carving technique described in Appendix~\ref{sec:Gauss_corr} on a spatial light modulator (SLM, Holoeye Pluto-2).
The photon is subsequently demagnified through a 4f-system and sent into the loop performing the unitary modulation. 
Within the loop, the lenses before and after the beamsplitter (BS) create a continuous string of 4f-systems which ensure that the beam has the same waist and wavefront curvature when exiting the loop, regardless of the number of loops.
Inside the loop, the half-wave plates (HWP) and Dove prism (Dove) perform the rotation of the transverse structure while keeping the polarization of the photon intact.
After the loop, the photon is imaged onto a separate side of the same SLM using a final magnifying 4f-system.
The SLM, in conjunction with the SMF coupling, performs the spatial mode filtering described in Appendix~\ref{sec:Gauss_corr}.
Finally a second SPAD is used to detect the signal photon, and a coincidence counter (CC: ID Quantique ID900) is used to post-select on photon pairs while measuring the delay between the arrival of the heralding and signal photon.

\section{Gaussian correction to spatial mode generation and measurement}
\label{sec:Gauss_corr}
One powerful method for generating and measuring spatial modes using a single phase-only SLM is to include the complex amplitude information in the diffraction grating of the hologram \cite{Bolduc,bouchard2018measuring,davis1999encoding}.
The method relies on encoding the wanted phase and amplitude structure onto the first diffraction order of a hologram, where the amplitude structure is encoded by spatially varying the diffraction efficiency of the grating.
Hence, although these methods can produce the wanted field very accurately, they are intrinsically lossy.

For the generation, the methods aim to shape an incident plane wave into a field with a transverse scalar field structure $\Psi(\vek{\rho}) = A(\vek{\rho})e^{\textrm{i}\Phi(\vek{\rho})}$, where $\vek{\rho} = (x,y)$ labels the transverse coordinates, $A(\vek{\rho})=|A(\vek{\rho})|$ is the normalized amplitude structure, and $\Phi(\vek{\rho})$ is the transverse phase structure of the field.
This shaping is done using a phase-only hologram with a phase structure \cite{Bolduc}
\begin{equation}
\label{eq:holo_phase}
     M(\vek{\rho}) \text{Mod} \left(F(\vek{\rho}) + \frac{2\pi x}{\Lambda_x},2\pi \right),
\end{equation}
where $\Lambda_x$ is the spatial frequency of the blazed grating which diffracts the field in the x-direction.
The functions $M$ and $F$ need to be chosen such that the hologram produces the correct field in the first diffraction order, i.e., $M(\vek{\rho}) =  1 + \sinc^{-1}(A(\vek{\rho}))/\pi$ and $F(\vek{\rho}) = \Phi(\vek{\rho})-\pi M(\vek{\rho})$ \cite{Bolduc}.
Here the M-function has values between $0\leq M \leq 1$ and $\sinc^{-1}$ is the inverse function of the unnormalized $\sinc(x)$-function which operates in the domain $x\in[-\pi,0]$, as explained in Ref.~\cite{Bolduc}.
After imprinting this phase on the incident plane wave, the resulting field in the first diffraction order can be calculated using a Taylor-Fourier expansion, giving the field \cite{Bolduc}
\begin{equation}
   T_1(\vek{\rho}) = -\sinc\left(\pi[M-1]\right)e^{\textrm{i}(F+\pi M)}. 
\end{equation}
Hence, we get the wanted field in the first diffraction order (up to a global phase) when the previously mentioned forms of the functions $M$ and $F$ are used.

However, in experiments the condition of an incident plane wave can only be met approximately, e.g., by using a large Gaussian beam and applying the hologram only to the center of this Gaussian.
While we still used a Gaussian input beam in the work presented here, we used a different method for getting rid of the Gaussian profile of the incident field.
Although the Gaussian beam was still slightly larger than the hologram, we did not modulate only the center part of it and instead got rid of this Gaussian field structure $\Psi_G(\vek{\rho};w_{in})$ by including the normalized Gaussian amplitude $A_G(\vek{\rho};w_{in})$ in M: 
\begin{equation}
\label{eq:Gauss_corr}
    M =  1 + \frac{1}{\pi}\sinc^{-1}\left(\frac{A(\vek{\rho})}{A_G(\vek{\rho};w_{in})}\right).
\end{equation}
In the experiment, we measured the radius $w_{in}$ of the input Gaussian and applied this correction to the hologram.
Hence, the incident Gaussian beam modulated by the phase-only hologram given in Eq.~\eqref{eq:holo_phase}, produces the wanted field in the first diffraction order.
Although this correction improved our data, the fidelity of the field produced by this method is still limited by the resolution of the SLM.

The same phase-only hologram method can be used to measure any transverse-spatial mode when the first diffraction order is coupled into an SMF after the hologram 
\cite{bouchard2018measuring}.
In the hologram, the complex conjugate of the measured transverse structure $H_{holo}(\vek{\rho})=\Psi_{meas}^*(\vek{\rho})$ is imprinted on the first diffraction order, and the probability of detecting a photon after the single mode fiber depends on the overlap of the normalized structure of the incident field $\Psi_{in}(\vek{\rho})$, the structure imprinted by the hologram, and the normalized Gaussian eigenmode of the SMF (backward propagated to the hologram) \cite{bouchard2018measuring}
\begin{equation}
    \int_0^{2\pi}\int_0^{r_{max}} r H_{holo}(\vek{\rho})\Psi_{in}(\vek{\rho})\Psi_G(\vek{\rho};w_{out})drd\varphi.
\end{equation}
Again, if the Gaussian in the above equation could be approximated as a plane wave, and the radial extent of our system aperture was infinite, the measurement outcome would only depend on the overlap integral between the incident field $\Psi_{in}$ and the mode structure imprinted by the hologram \cite{bouchard2018measuring}.
However, since filtering for a single plane wave is not possible, the Gaussian mode of an SMF needs to be removed from the equation.
One option is again to enlarge the size of the backwards propagated Gaussian as is done in the reference \cite{bouchard2018measuring}.
But similar to the mode generation, by including the Gaussian amplitude of the SMF into our hologram, its effects can be minimized from the overlap integral while only requiring that the backwards propagated Gaussian is slightly larger than the incident mode
\begin{equation}
    H_{holo} = \frac{\Psi_{meas}^*(\vek{\rho})}{A_G(\vek{\rho};w_{out})}.
\end{equation}
This correction can be implemented with the same change in the $M$ function that is introduced in equation~\eqref{eq:Gauss_corr}.

\section{Precise alignment of the Dove prism angle}
\label{sec:Dove_angle}
To be confident that our QFA performed the correct operation with each cycle, we developed a method to improve the precision of the rotation angle of the Dove prism, which is limited to $\pm\ang{0.5}$ by the optical mount.
We noted that for a given prism angle $\phi$, a petal mode as defined in Eq. \eqref{eq:petal_basis} with OAM $\ell$ can be generated with rotational symmetry for all loops $n$ according to the relations $\ell=\ang{360}/(2\phi)$ or $\ell=\ang{360}/(4\phi)$ for odd or even OAM values, respectively.
We therefore displayed identical petal-mode holograms on both the generation and measurement screens of the SLM, then recorded the power over time signal of a $\lambda=810$~nm alignment laser while rotating the measurement hologram's structure in $\ang{0.1}$ increments.

\begin{figure}[h]
    \centering
    \includegraphics[width=0.8\textwidth]{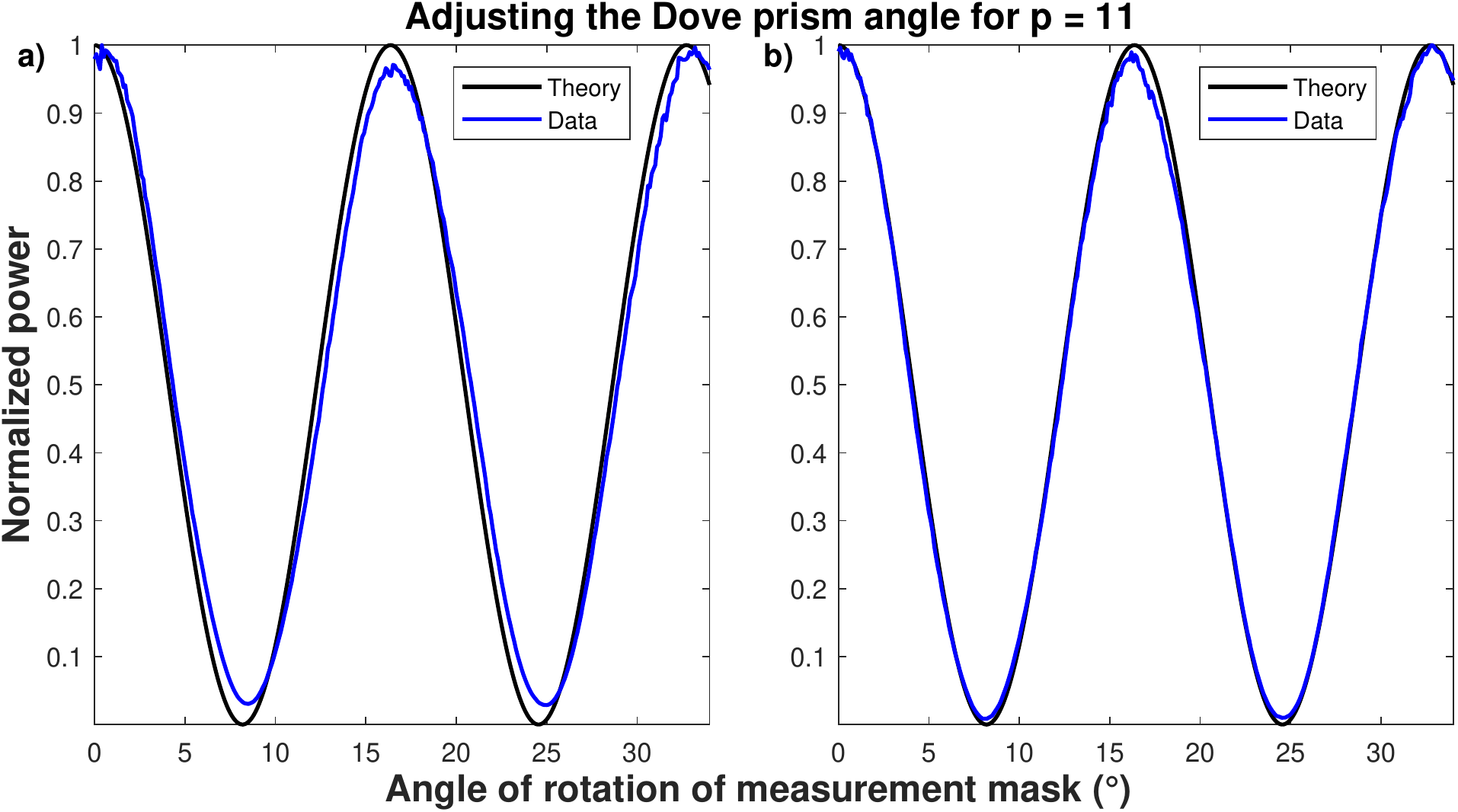}
    \caption{Laser power over rotation angle of the measurement hologram for \textbf{a)} the third iteration of Dove angle alignment and \textbf{b)} the eighth iteration. The data were recorded for rotational offsets of the measurement mask hologram for every $\ang{0.1}$ from $\ang{0}$ to $\ang{34}$. The data in \textbf{a)} deviate by ${\sim}\ang{0.5}$ from the theoretical curve, indicating that the Dove prism was under-rotated by ${\sim}\ang{0.25}$ (an over-rotation would cause the signal to drop too quickly). A sinusoidal fit of the data in \textbf{b)} reported an offset of ${\sim}\ang{0.01}$.}
    \label{fig:Dove_angle_check}
\end{figure}

As shown in Fig.~\ref{fig:Dove_angle_check}, this process results in a sinusoidal power reading as a function of mask angle.
If an imperfection exists in the overlap of the petal mode, it is magnified with each loop. 
The degree of deviation of the minima of the signal from the minima of the theoretical curve $\text{cos}^2(\ell\theta)$ dictates the degree of over- or under-rotation of the Dove prism angle (to a factor of $1/2$).

After performing the measurement and adjusting the angle $\phi$ accordingly, the procedure was repeated.
We iterated this process until the data agreed with the theoretical curve to better than $\ang{0.1}$.
Once confident that the prism angle did not contribute significantly to operational errors, we performed the QST and QFA measurements reported in Sec.~\ref{sec:results}.

\section{Gauging of the loop-dependent efficiency and accepting probabilities}
\label{sec:loops_acc}
We obtained the accepting probabilities reported in Fig.~\ref{fig:QFA_data1} and Fig.~\ref{fig:QFA_2cycle} by comparing the time-domain histogram of the coincidence counts of a Gaussian mode with that of a QFA mode.

\begin{figure}[h]
    \centering
    \includegraphics[width=0.7\textwidth]{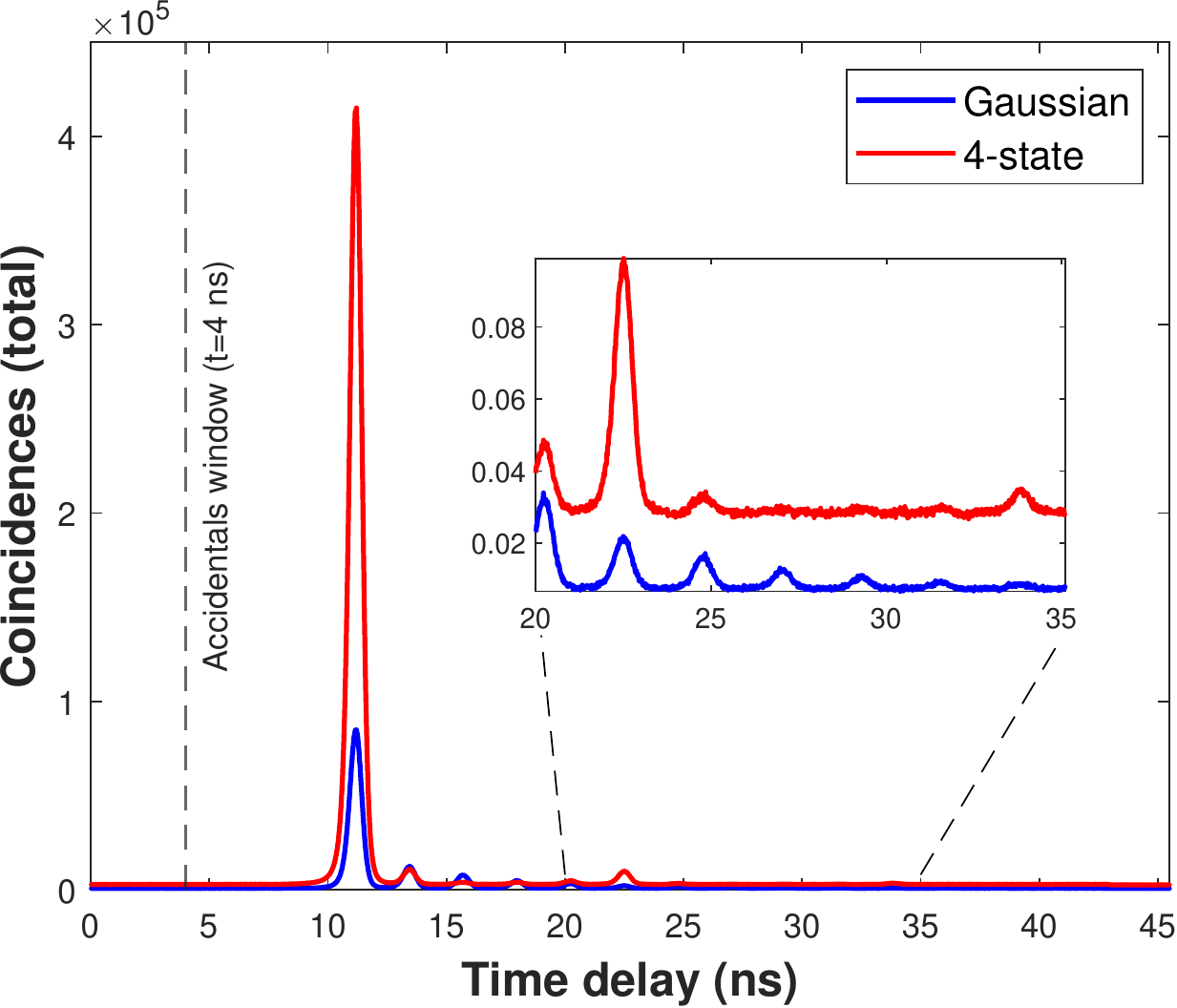}
    \caption{Example coincidence histograms for the Gaussian state (blue) and the $4$-state QFA with OAM values of $\ell=\{1,3\}$ (red) recognizing the $MOD_5$ language for a string of length $10$. The highest red peak in the zoomed plot corresponds to the $5^{\text{th}}$ loop, and the last peak represents the $10^{\text{th}}$. The data shown is the sum over all collected histograms, i.e., a measurement time of $2000$~seconds for the Gaussian and $26000$~seconds for the 4-state QFA, that contribute to the statistical data reported in Fig.~\ref{fig:QFA_2cycle}~\textbf{a)}. The $70$:$30$ beamsplitter was used with the Dove prism at $\ang{18}$.}
    \label{fig:hist}
\end{figure}

As shown in Fig.~\ref{fig:hist}, the histograms contain peaks in the signal where significant numbers of coincidences are detected.
Because the loop is roughly 60~cm long, each operation performed on a signal photon delays it by around two nanoseconds, so the peaks have a $\Delta t\approx2$~ns delay between them.
If we detect the zeroth loop at time $t_0$, then the $n^{\text{th}}$ loop is detected at approximately $t_0 + n \Delta t$~ns.
To calculate the number of coincidences in the $n^{\text{th}}$ loop, we defined a ${\sim}1$~ns window and summed over the counts in all the time bins contained in the $n^{\text{th}}$ peak.
With the Time Controller software of the ID900 in high resolution mode, each time bin is $\tau=13$~ps, so we summed over $77$ bins per loop.
We centered each window around the maximum of each peak of the Gaussian data after dividing the total measurement time into ${\sim}50$ datasets, summed and averaged.
The delay between peaks was found to be $174$ bins, or ${\sim}2.26$~ns. 
Note that the delay time between the peaks was chosen to be the same for all recorded peaks and measurements. 

To get a true measure of the coincidence counts, we needed to account for accidental coincidences, i.e., simultaneous detection events at both detectors (within the coincidence window $\tau$) that are not caused by actual photon pairs but dark counts of the detectors, detection of background photons, or the detection of photons from uncorrelated pairs \cite{ramelow2013highly}. 
The number of accidental coincidences can be estimated to be $R_1 R_2 \tau$, where $R_j$ is the rate of single photons at detector $j$.
Typical values of $R_j$ were ${\sim}10^6$~Hz in the heralding arm and between ${\sim}\num{1.5e3}$~Hz and ${\sim}\num{3.0e4}$~Hz in the signal arm, where the upper end is reached when sending a Gaussian through the QFA setup. 
On the other hand, when generating higher-dimensional QFA states more light is carved away at the SLM, thus leading to lower detection rates.
However, instead of estimating the accidental coincidences using the above mentioned formula, we determined the experimental rate of accidentals by evaluating the coincidence detections at an electronic delay setting where no coinciding photons from the source should be observed, i.e., we started to record the coincidence histogram approximately 10~ns before the actual arrival of the signal photon and averaged the measured accidental coincidences over the first 4~ns (see Fig. \ref{fig:hist}).
The obtained constant rate of accidental coincidence detections was then subtracted from the measured coincidences.

If we define the coincidence counts per loop as $C_n$ and the average accidentals per bin $b$ as $\bar{A}$, then we can write the counts per loop for mode $M$ as
\begin{equation}
    C_n^M = \sum_{b=1}^{b=77}(C_b^M - \bar{A}).
    \label{eq:Cn}
\end{equation}
We obtained the Gaussian mode counts, $C_n^G$, and the QFA mode counts, $C_n^Q$, then calculated the accepting probability as follows: first, we normalized the zeroth loop of each dataset of mode $M$ to 1 to get the relative accepting probability of the $n^{\text{th}}$ loop of mode $M$, $P_n^M = C_{n}^M/C_0^M$.
Then we found the average probabilities of the datasets, $\bar{P}_n^M$.
Lastly, we normalized the probability of the QFA mode to the Gaussian counts:
\begin{equation}
    P_n=\frac{\bar{P}_n^Q}{\bar{P}_n^G}.
    \label{eq:prob_counts}
\end{equation}
By this process we arrived at the accepting probability of the QFA state $V_{\cent}\ket{v_0}$ described by the overlap amplitude of Eqs.~\eqref{eq:probability} and \eqref{eq:acc_prob_QFA}.

\section{Quantum state tomography}
\label{sec:QST_app}
\begin{figure}[htb]
    \centering
    \includegraphics[width=0.6\textwidth]{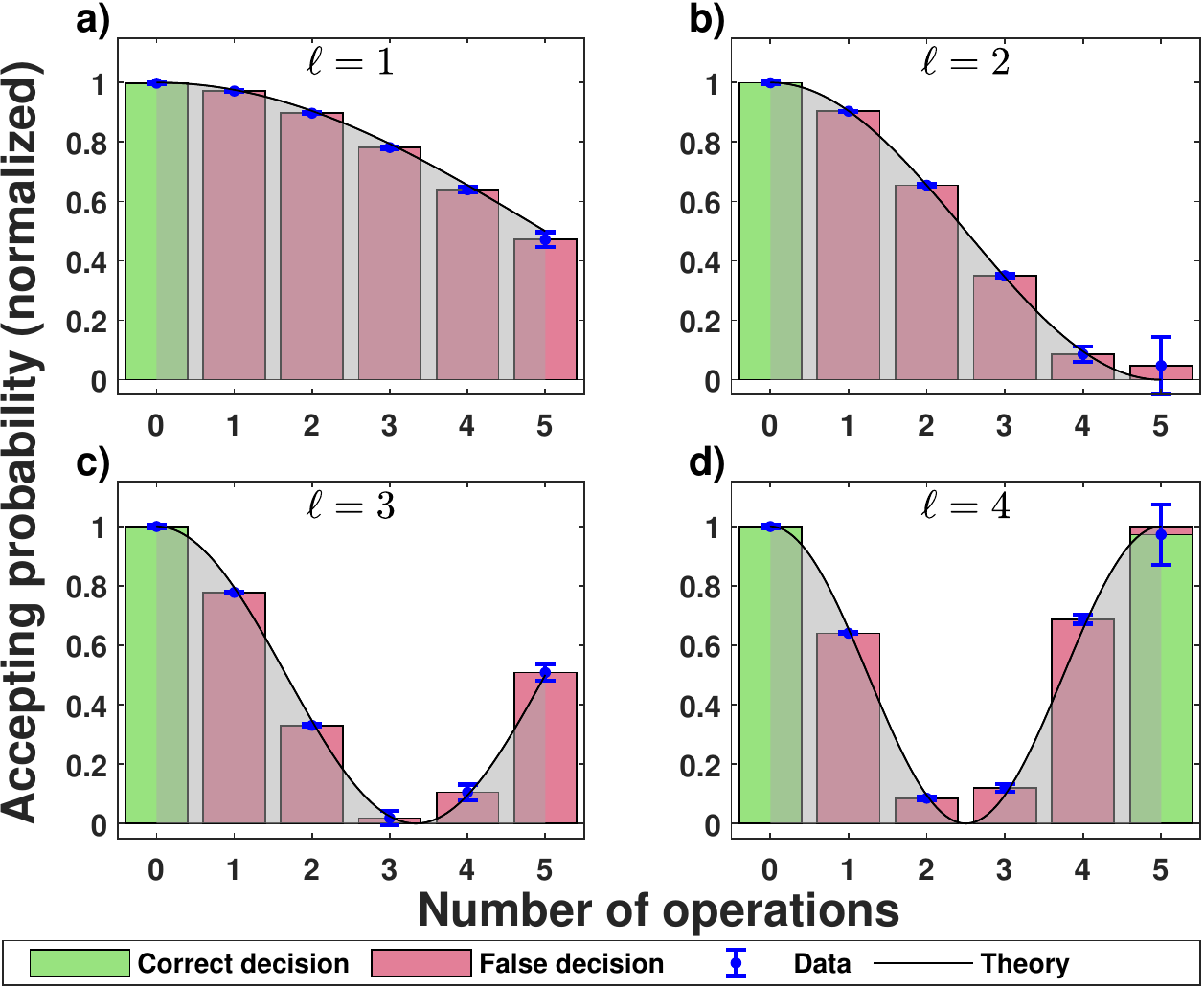}
    \caption{Measurements of the QST states of Fig.~\ref{fig:QST} converted to accepting probabilities. Note that the frequency of the sinusoidal probability curve increases with increasing OAM $\ell$. For \textbf{a)}, $\ell=1$, the accepting probability remains high for the first several operations, ending near $50$~\%. For \textbf{b)}, $\ell=2$, the final state is orthogonal to the initial state. For \textbf{c)}, $\ell=3$, the final state returns to around $50$~\%, and for \textbf{d)}, $\ell=4$, the system performs as a $2$-state QFA recognizing $MOD_5$. The Dove prism was at $\ang{4.5}$ with a $50$:$50$ beamsplitter. The error bars represent $\pm1$ standard deviation.}
    \label{fig:QST_greenbars}
\end{figure}

Here we show more details about the QST measurements. 
In Fig.~\ref{fig:QST_greenbars}, we see that the state vectors on the Bloch sphere of Fig.~\ref{fig:QST} can be represented in terms of accepting probabilities.
With the Dove prism at $\ang{4.5}$, the petal mode with $\ell=4$ acts as a $2$-state QFA recognizing $MOD_5$.
To demonstrate how many photons contributed to the quantum state tomography, we show the total heralded single photon events (coincidence counts) for each projective measurement and per single repetition in Table~\ref{tab:QST}. 
We repeated each measurement 30 - 70 times and used the data to perform a state reconstruction via direct inversion \cite{Schmied2016QST,Chithrabhanu2015generalized}.

\begin{table}[htb]
 \begin{tabular}{ || c || c | c | c | c | c | c || }
 \hline
 \multicolumn{1}{||c||}{Input state} & \multicolumn{6}{c||}{Photon number ($\times 10^5$) per projected state} \\ 
 \hline
 \ & $\ket{z^+}$  & $\ket{z^-}$ & $\ket{x^+}$ & $\ket{x^-}$ & $\ket{y^+}$ & $\ket{y^-}$ \\
 \hline\hline
 $\ket{p^+_{\ell_1}}$ & 2.00 & 2.28 & 2.15 & 0.09 & 1.39 & 0.87 \\ 
 \hline
 $\ket{p^+_{\ell_2}}$ & 2.21 & 2.04 & 1.98 & 0.26 & 1.49 & 0.77 \\
 \hline
 $\ket{p^+_{\ell_3}}$ & 1.80 & 1.80 & 1.54 & 0.38 & 1.31 & 0.62 \\
 \hline
 $\ket{p^+_{\ell_4}}$ & 1.36 & 1.37 & 1.09 & 0.36 & 0.96 & 0.50 \\ [1ex]
 \hline

\end{tabular}
  \caption{Photons per repetition for the QST measurements of Fig.~\ref{fig:QST} and Fig.~\ref{fig:QST_greenbars}. The projected states were $\ket{z^{\pm}} = \ket{\pm\ell}$, $\ket{x^{\pm}} =\frac{1}{\sqrt{2}}(\ket{\ell}{\pm} \ket{-\ell})$, and $\ket{y^{\pm}}=\frac{1}{\sqrt{2}}(\ket{\ell}{\pm}i\ket{-\ell})$. These values are based on summing the total heralded photons, for all six projections, from the zeroth through fifth Dove prism operations, then dividing by the number of repetitions ($30$ to $70$) of the measurement. We normalized to Gaussian counts based on $30$ repeated measurements of around $\num{2.18e5}$ photons per repetition.}
  \label{tab:QST}

\end{table}

\section{Higher-dimensional ($2d>p$) data}
\label{sec:high-d}

\begin{figure}[htb]
    \centering
    \includegraphics[width=0.6\textwidth]{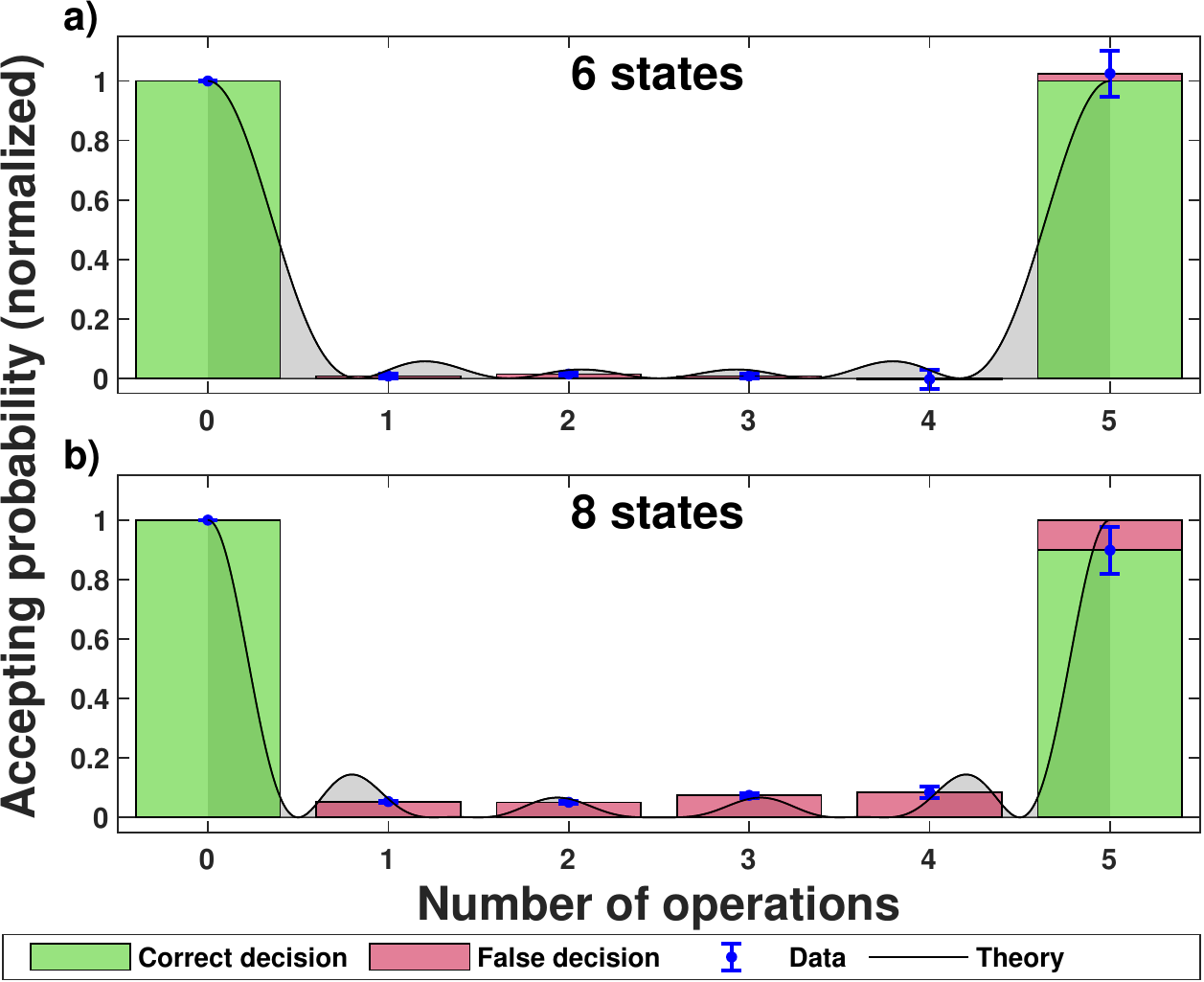}
    \caption{Higher-dimensional OAM states recognizing $MOD_5$ drive the false decision probability down. In \textbf{a)}, the Dove prism was at $\ang{18}$, and we utilized Eq.~\eqref{eq:petals_initial_state} with $\ell=\{1,3,5\}$. The data (blue points) and error bars ($\pm1$ standard deviation) were obtained from $90$ repetitions of the experiment with around $\num{3.56e4}$ heralded photons per run. In \textbf{b)}, the Dove prism was at $\ang{36}$, and we superposed $\ell=\{1,2,3,4\}$. The Time Controller software was not in high-resolution mode for this measurement, so the bin width was $100$~ps rather than $13$~ps ($10$ bins per loop). The data and error bars (blue points) were obtained from $77$ repetitions of the experiment with around $\num{3.32e4}$ heralded photons per run. The theoretical curves for continuously rotating states were calculated using Eq.~\eqref{eq:acc_prob_QFA}.}
    \label{fig:QFA_6_8}
\end{figure}

To further illustrate how photon OAM can be used to construct higher-dimensional QFA states that reduce the false accepting probability, we include data that recognize $MOD_5$ while not demonstrating a quantum advantage.
As shown in Fig.~\ref{fig:QFA_6_8}~a)-b), we measured data for both a 6-state QFA and an 8-state QFA.

We used petal modes with OAM values of $\ell=\{1,3,5\}$ to construct the 6-state QFA, which led to false decisions of $1.42\pm 0.40~\%$ (loop 2) or less with an accepting probability of $102.50\pm 7.71~\%$ after 5 operations.
The 8-state QFA with OAM values of $\ell=\{1,2,3,4\}$ resulted in decision errors of $8.42\pm 1.89~\%$ (loop 4) or less, and the final accepting probability was $89.87\pm 7.82~\%$ for inputs of length 5.

\end{document}